\documentclass[prd,aps,floats,preprintnumbers,preprint]{revtex4}

\usepackage{graphicx}
 
\newcommand{\be}{\begin{equation}}
\newcommand{\ee}{\end{equation}}
\newcommand{\bea}{\begin{eqnarray}}
\newcommand{\eea}{\end{eqnarray}}

\begin{document}


\setlength{\unitlength}{1mm}
\title{Low red-shift formula for the luminosity distance in a LTB model with cosmological constant}
\author{Antonio Enea Romano$^{1,3,4}$}
\email{aer@phys.ntu.edu.tw}
\author{Pisin Chen$^{1,2}$}
\email{pisinchen@phys.ntu.edu.tw}
\affiliation
{
${}^1$Leung Center for Cosmology and Particle Astrophysics, National Taiwan University \\
${}^2$Kavli Institute for Particle Astrophysics and Cosmology, SLAC National Accelerator Laboratory, Menlo Park, CA 94025, U.S.A. \\
$^3$Instituto de Fisica, Universidad de Antioquia, A.A.1226, Medellin, Colombia\\
$^4$Yukawa Institute for Theoretical Physics, Kyoto University,
Kyoto 606-8502, Japan
}
\begin{abstract}
We calculate the low red-shift Taylor expansion for the luminosity distance for an observer at the center of a spherically symmetric matter inhomogeneity with a non vanishing cosmological constant.  
We then test the accuracy of the formulas comparing them to the numerical calculation for different cases for both the luminosity distance and the radial coordinate.
The formulas can be used as a starting point to understand the general non linear effects of a local inhomogeneity in presence of a cosmological constant, without making any special assumption about the inhomogeneity profile.
\end{abstract}
\maketitle
\section{Introduction}
Modern cosmological observations such as the luminosity distance  \cite{Perlmutter:1999np,Riess:1998cb,Tonry:2003zg,Knop:2003iy,Barris:2003dq,Riess:2004nr} and
the WMAP measurements \cite{WMAP2003,Spergel:2006hy} of cosmic
microwave background radiation (CMBR)  have provided a strong evidence for the presence of dark energy.
One of the main assumptions of the standard cosmological model used in fitting these observational data is spatial homogeneity of the Universe. We cannot nevertheless exclude the presence of a local inhomogeneity around us which could affect our interpretation of cosmological data \cite{Romano:2010nc,Sinclair:2010sb,Romano:2011mx}.

So far most of the efforts in estimating these effects have consisted in using some ansatz for the profile of the inhomogeneity and then calculate numerically the effects on cosmological observables.
Such an approach has the limitation of depending on the particular functional form chosen to model the local inhomogeneity, and of relying completely on numerical calculations.
In order to provide a more general study of this effects we approach the problem analytically and we derive a low-redshift formula for the luminosity distance relation for an observer at the center of a matter inhomogeneity in presence of a cosmological constant modeled by a LTB solution. 

The paper is organized as follows
We first calculate the low red-shift expansion of the null radial geodesics for a central observer and then use it to obtain the luminosity distance. The calculation is based on using the analytical solution, and the geodesic equation expressed in the same coordinates of the analytical solution.
The formula obtained is then compared to the numerical calculation of the luminosity distance to test its accuracy.
In the appendix we give details of the derivation and the simplified formulae in the limit in which the inhomogeneity can be treated perturbatively.
\section{LTB solution with a cosmological constant\label{ltb}}
The LTB solution can be expressed in the form
as \cite{Lemaitre:1933qe,Tolman:1934za,Bondi:1947av} as
\begin{eqnarray}
\label{LTBmetric} %
ds^2 = -dt^2  + \frac{(R,_{r})^2 dr^2}{1 + 2\,E(r)}+R^2
d\Omega^2 \, ,
\end{eqnarray}
where $R$ is a function of the time coordinate $t$ and the radial
coordinate $r$, $E(r)$ is an arbitrary function of $r$, and
$R_{,r}=\partial_rR(t,r)$.
We can get from the Einstein's field equations 
\begin{eqnarray}
\label{eq2} ({\frac{\dot{R}}{R}})^2&=&\frac{2
E(r)}{R^2}+\frac{2M(r)}{R^3}+\frac{\Lambda}{3} \, , \\
\label{eq3} \rho(t,r)&=&\frac{2M,_{r}}{R^2 R,_{r}} \, ,
\end{eqnarray}
where $M(r)$ is an arbitrary function of $r$ which arises in the integration of one of the Einstein's equations respect to time, $\dot
R=\partial_tR(t,r)$ and we are assuming $c=8\pi G=1$. 

The derivation of the analytical solution \cite{Dilwyn} is based on the introduction of a new coordinate $\eta=\eta(t,r)$ and a variable $a$ by 
\bea
(\frac{\partial\eta}{\partial
t})_r=\frac{r}{R}\equiv\frac{1}{a}\,, \label{etadef} 
\eea
and new functions by 
\bea
\rho_0(r)\equiv\frac{6 M(r)}{r^3}\,,\quad
k(r)\equiv-\frac{2E(r)}{r^2}\,. \eea 
We can then express  Eq. (\ref{eq2}) in the form
\be 
(\frac{\partial a}{\partial\eta})^2 =-k(r)
a^2+\frac{\rho_0(r)}{3} a+\frac{\Lambda}{3} a^4\,, 
\ee 
where $a$ is now
a function of $\eta$ and $r$, $a=a(\eta,r)$. 
The coordinate $\eta$, which can be considered a generalization of
the conformal time in a homogeneous FLRW universe, is defined 
implicitly  by Eq.~(\ref{etadef}). The  relation
between $t$ and $\eta$ is
\be
t(\eta,r)=\int_0^{\eta}{a(x,r)dx}+t_b(r) \,,
\ee
which can be computed analytically, and involve elliptic integrals of the third kind\cite{ellint}.

The function $t_b(r)$ is a constant of integration, , also called bang function, since by at time $t=t_b(r)$ we have $a(t_b(r),r)=0$. This corresponds to the possibility that the big bang can happen at different times at different positions from the center in a LTB space. 
Its gradient is related to the decaying modes of early universe density perturbation, and which CMB observations strongly constraint to be small.  
In the rest of this paper we will consider a homogeneous big bang, i.e. we will have
\be
t_b(r)=0
\ee
which in terms of early universe cosmological perturbations corresponds to growing modes only.
The solution is given by:
\be
a(\eta,r)
=\frac{\rho_0(r)}{3\phi\left(\frac{\eta}{2 };g_2(r),g_3(r)\right)+k(r)}\,,
\ee
where $\phi(x;g_2,g_3)$ is the Weierstrass elliptic function
which satisfies the differential equation 
\be
\left(\frac{d\phi}{dx}\right)^2=4\phi^3-g_2\phi-g_3\, ,\label{eqweir} 
\ee 
and
\bea
g_2=\frac{4}{3}k(r)^2 \,,\quad
g_3=\frac{4}{27} \left(2k(r)^3 -\Lambda\rho_0(r)^2\right)\,.
\eea
In this paper we will choose the so called FLRW gauge, i.e. the coordinate system in which $\rho_0(r)$ is constant.
It is convenient to write the solution in terms of dimensionless quantities 
\cite{EneaRomano:2011aa}:

\bea
k(r)&=&(a_0 H_0)^2 K(r)\,, \\
\eta&=&T(a_0 H_0)^{-1}\,,\\
\rho_0(r)&=&3 \Omega^0_M(r) a_0^3 H_0^2\,,\\
\Lambda&=&3 \Omega_{\Lambda}H_0^2\,, \\
a(\eta,r)&=&a(T(a_0 H_0)^{-1},r)=\tilde{a}(T,r) \,, \\
\eea
to obtain
\bea
\tilde{a}(T,r)&=&\frac{3 a_0\Omega^0_M(r)}{K(r)+12 \tilde{\phi} (T,g_2(r),g_3(r))} \,,\\
g_2(r)&=&\frac{K(r)^2}{12} \,,\\
g_3(r)&=& \frac{1}{432} (2 K(r)^3-27 \Omega_{\Lambda} (\Omega^0_M(r))^2)\,.
\eea
We can relate the solution expressed in the two different forms by multiplying every term by $(a_0 H_0)^2$ and using the original dimensionful quantities $\eta,k(r),\rho_0(r)$
\bea
a(\eta,r)&=&\frac{\rho_0(r)}{k(r)+12 \phi (\eta,g_2(r),g_3(r))}=\tilde{a}(T,r) \,,\\
\phi(\eta,r)&=&\tilde{\phi}(\eta (a_0 H_0),r)(a_0 H_0)^2=\tilde{\phi}(T,r)(a_0 H_0)^2\,. 
\eea
In this form $H_0$ is an arbitrary scale which we can set equal to the observed value, which will also coincide with the $H_0^{LTB}$ by appropriately setting the value of $T_0$ as explained in more details in  \cite{EneaRomano:2011aa}.
Without any loss of generality we can choose a coordinate system in which $\rho_0(r)=const.$, implying that $\Omega^0_M(r)=const.$ , which we will simply denote as $\Omega_M$ in the rest of the paper.
\section {Geodesic equations and luminosity distance}
We will
solve \cite{Romano:2009xw} the null geodesic equation written in terms of the
coordinates $(\eta,r)$. We then perform a local expansion of the
solution around $z=0$ corresponding to the point $(t_0,0)$, or
equivalently $(\eta_0,0)$, where $t_0=t(\eta_0,0)$. 

The luminosity distance for an observer located at the center of a  LTB
space-time is given by  
\be
D_L(z)=(1+z)^2 R(t(z),r(z)) =(1+z)^2
r(z)a(\eta(z),r(z)) \,, \ee 
where
$\Bigl(t(z),r(z)\Bigr)$ or $\Bigl((\eta(z),r(z)\Bigr)$ is the
solution of the null radial geodesic equations as a function of $z$. 
The equation for geodesics can be easily obtained in the coordinates $(t,r)$ 
\bea \label{geo1}
\frac{dt}{dr}
=-\frac{R_{,r}(t,r)}{\sqrt{1+2E(r)}} \,. \eea
where $t=T(r)$ is the time coordinate along the light-like radial geodesic as a
function of the coordinate $r$. 
Using the definition of redshift and
by following the evolution of a short time interval along the null
geodesic $T(r)$, from Eq. (\ref{geo1}) we get 
 \cite{Celerier:1999hp}:
\begin{eqnarray}
{dr\over dz}&=&{\sqrt{1+2E(r(z))}\over {(1+z){\dot R}_{,r}[r(z),t(z)]}} \,,
\nonumber\\
{dt\over dz}&=&-{R_{,r}[r(z),t(z)]\over {(1+z){\dot
R}_{,r}[r(z),t(z)]}} \,. \label{eq:35}
\end{eqnarray}
We can now \cite{Romano:2009xw} express the above geodesics equations in the coordinates $(\eta,r)$ : 
\bea \label{geo3}
\frac{d \eta}{dz} &=&-\frac{\partial_r
t(\eta,r)+F(\eta,r)}{(1+z)\partial_{\eta}F(\eta,r)}
\equiv p(\eta,r) \,,\\
\label{geo4} \frac{dr}{dz}
&=&\frac{a(\eta,r)}{(1+z)\partial_{\eta}F(\eta,r)} \equiv
q(\eta,r) \,, \eea where \be F(\eta,r)\equiv \frac{\
R_{,r}}{\sqrt{1+2E(r)}}=
\frac{1}{\sqrt{1-k(r)r^2}}[\partial_r (a(\eta,r) r)
-a^{-1}\partial_{\eta} (a(\eta,r) r)\, \partial_r t(\eta,r)]
\,. \ee
where the functions $p,q,F$ have explicit
analytical forms, making it particularly suitable to derive analytical results.
\section {Formula for the luminosity distance}
In order to obtain the redshift expansion of the luminosity distance we need to expand the relevant functions: 
\bea
k(r)&=&(a_0 H_0)^2 K(r)=K_0+K_1 r+K_2 r^2+ ..
\eea
We will use eq.(7) to obtain the expansion for $t(\eta,r)$ from the exact solution for $a(\eta,r)$. 

After integration we  obtain:
\bea
t(\eta,r)&=&\int_0^{\text{$\eta_0 $}} a(x,r) \, dx+
   a(\eta_0,r)(\eta -\eta_0)+\frac{1}{2} 
   a'(\eta_0,r)(\eta -\eta_0)^2+\frac{1}{6} 
   a''(\eta_0,r)(\eta -\eta_0)^3+.
\eea
Using the expression above we can obtain the expansion of $t(\eta,r)$ directly from the expansion of $a(\eta,r)$ except for the first term which involves the integral of an elliptic function.
The expansion respect to the radial coordinate $r$ is straightforward and we will not report here all the intermediates results but only the final expression for the solution of the geodesics equations.
In this paper we provide the first derivation of the expansion of $t(\eta,r)$ while in previous works the coefficients were not evaluated explicitly in terms of $K_i$.
As a consequence the formulae we obtain only depend on $K_i$, and do not require any addition calculation. 

We can now find a local Taylor expansion in red-shift for the geodesics equations  \cite{Romano:2010nc}, and then calculate the luminosity distance.
The general expression is rather cumbersome, and is given in the appendix.He re we will report only the result assuming $K_0=0$, which is still showing the general nature of the effect.
From a physical point of view fixing $K_0$ does not affect the value of $H_0$, but it does affect the age of the Universe as shown in \cite{EneaRomano:2011aa}, but using the freedom in the choice of the bang function it is possible to any obtain any age, by appropriately fixing it to a constant value $t_b(r)=t_0$, while since $t'_b(r)=0$ there would not be any problem related with the compatibility with early universe perturbations which should not contain decaying modes.

We will expand the solution of the geodesic equations according to:
\bea
r(z)&=&r_1 z+r_2 z^2 +... \,\\
\eta(z)&=&\eta_0+\eta_1 z+\eta_2 z^2 +... \,\\
K(z)&=&K_1 z+K_2 z^2 +... \,
\eea
After substituting in the geodesics equation we can map the solution of the system of differential equations into a system of algebraic equations for the coefficients of the above expansions.
The general expression is rather long and complicated, so here we will report the much simpler case when $K_0=0$, while in the appendix we give more general formulae.
The motivation for considering the $K_0$ case is to focus on the effects of the inhomogeneities which are captured by ${K_1,K_2}$, while $K_0$ corresponds to the homogeneous component of the curvature function, which in absence of inhomogeneities is simply the curvature of a FLRW model, and as such is not associated to any new physical effect not already known from standard cosmology.
 
For the geodesics we get:
\bea
\eta_1 &=&-\frac{K_1 (T_0-1) T_0+3 \Omega_M}{3 a_0 H_0 \Omega_M} \,,\\ \nonumber
\eta_2 &=& \frac{1}{36 a_0 H_0 \Omega_{\Lambda} \Omega_{M}^2)}\bigg[3 \Omega_{\Lambda} \Omega_{M} (9 \Omega_{M}^2 - 4 K_2 (-1 + T_0) T_0) + 3 K_1 \Omega_{\Lambda} \Omega_{M} (-4 + (4 - 9 \Omega_{M}) T_0 + \\ \nonumber
  &&+ (-4 + 9 \Omega_{M}) T_0^2) + 
  K_1^2 T_0 (2 \Omega_{\Lambda} (2 + (-4 + 3 \Omega_{M}) T_0 - 6 (-1 + \Omega_{M}) T_0^2 + 
        3 (-1 + \Omega_{M}) T_0^3) \\ 
        &&- 4 (-1 + T_0) WZ + \Omega_{M} (-4 + T_0 + 3 T_0 WZ))\bigg] \,,\\
r_1 &=& \frac{1}{a_0 H_0} \,,\\
r_2 &=& -\frac{1}{12 a_0 H_0 \Omega_{M}}\bigg[(9 \Omega_{M}^2 +  K_1(-4 + (4 - 6 \Omega_{M}) T_0 + (-4 + 6 \Omega_{M}) T_0^2)\bigg]\,,\\ \nonumber
r_3 &=& \frac{K_1^2}{{72 a_0 H_0 \Omega_{\Lambda} \Omega_M^2}}\bigg[2 \Omega_{\Lambda} (6 (3 \Omega_M^2-4 \Omega_M+1) T_0^4-12 (3 \Omega_M^2-4\Omega_M
   +1) T_0^3+3 (6 \Omega_M^2-13 \Omega_M+4) T_0^2 + \\ \nonumber
   & &+ 2 (9 \Omega_M-4)T_0+4)
   +3 \Omega_M^2 T_0 (3 T_0 \zeta_0+T_0-4) -2 \Omega_M (9 T_0^2 \zeta_0+T_0^2-6 T_0 \zeta_0-4         T_0+ \\ \nonumber
   & &6 \zeta_0-2)+8 (T_0^2-T_0+1)\zeta_0)+12 K_1 \Omega_{\Lambda} \Omega_M^2 ((9 \Omega_M-8) 		         T_0^2+(8-9 \Omega_M) T_0-5)+ \\ 
   & &+3\Omega_{\Lambda} \Omega_M (K_2 ((8-12 \Omega_M) T_0^2+4 (3 \Omega_M-2) T_0+8)+3 (9
    \Omega_M-4) \Omega_M^2) )\bigg] \,,
\eea
where 
\bea
T_0{(a_0 H_0)}^{-1}&=&\eta_0 \,,\\
\zeta_0&=&\zeta(\eta_0,g_2(0),g_3(0))\,,
\eea
and $\zeta$ is the Weierstrass Zeta Function satisfying the equation
\bea
\frac{d \zeta(z,g_2(r),g_3(r))}{d z}&=&-\phi(z,g_2(r),g_3(r)) \,.
\eea
The presence of this last function in the formulae obtained above is due to the fact that the function $t(\eta,r)$ which enters the geodesics equation is the integral of $a(\eta,r)$, and since this depends on $\phi(z)$, its integral will depend on $\zeta(z)$.
In the case of a LTB solution without a cosmological constant this integral can be performed without the introduction of a new function, while in this case it requires the introduction of $\zeta_0$ in the final formula.
 
The procedure to reduce the analytical formula to this form is rather complicated since it involves to express wherever possible all the intermediate expressions in terms of physically meaningful quantities and to use the properties of the elliptic functions. 
We give more details about it in the appendix.  
We can see that the effects of inhomogeneities start to show at first and second order respectively for $\eta(z)$ and $r(z)$. Contrary to the vanishing cosmological constant case, $T_0$ is now appearing explicitly in the formula. This is due to the fact in a LTB  model without cosmological constant is possible to express explicitly $T_0$ in terms of  $K_0$ and $q_0$, the central value of the deceleration parameter, while in our case we have:
\bea
q_0&=&-\frac{\ddot{a}(t_0,0)a(t_0,0)}{\dot{a}(t_0,0)^2} \,, \nonumber \\
   &=&\frac{3 \Omega_M}{2}-K_0-1  \,, \nonumber \\
   &=&-\frac{a_0^6 H_0^6 \left(K_0^3-54 \Omega_{\Lambda} \Omega_M^2\right)+9 a_0^4 H_0^4 K_0^2 \phi_0 +27 a_0^2 H_0^2 K_0 \phi_0 ^2+27 \phi_0 ^3}{2 a_0^6 H_0^6 \left(2 K_0^3-27 \Omega_{\Lambda} \Omega_M^2\right)+18 a_0^4 H_0^4 K_0^2 \phi_0 -54 \phi_0 ^3} \,,
\eea
where we have used the relations reported in the appendix to simplify the expression, and
\bea
t_0&=&t(\eta_0,0) \,, \\
\phi_0&=&\phi(\eta_0,g_2(0),g_3(0)) \,.
\eea
Such a relation constraint implicitly the value of $\phi_0$ in terms of cosmological parameters but it is not very useful to determine explicitly $T_0$, since it would involve to solve a cubic equation first and then to apply the inverse of an elliptic function, while in the vanishing cosmological constant case there exist a simple analytical relation because the Weierstrass function reduces to a trigonometric expression as shown in \cite{EneaRomano:2011aa}.

After substituting in the formula for the luminosity distance and expanding we finally get:
\bea
D^{\Lambda LTB}_L(z)&=&(1+z)^2r(z)a^{\Lambda LTB}(\eta(z),r(z))=D^{\Lambda LTB}_1 z+D^{\Lambda LTB}_2 z^2+D^{\Lambda LTB}_3 z^3 + . .\\
D^{\Lambda LTB}_1&=&\frac{1}{H_0} \,,\\
D^{\Lambda LTB}_2&=&-\frac{1}{4 H_0}(-4 + 3 \Omega_{M} + 2 K_1 (-1 + T_0) T_0) \,,\\
D^{\Lambda LTB}_3&=&\frac{1}{24
   H_0 \Omega_{\Lambda} \Omega_M}\bigg[K_1^2 (2 \Omega_{\Lambda} T_0 ((6 \Omega_M-5) T_0^3-2 (6 \Omega_M-5) T_0^2+6 (\Omega_M-1)
   T_0+2)+ \nonumber \\
   &&+\Omega_M T_0 (3 T_0 \zeta_0+T_0-4)-4 (T_0^2 \zeta_0-T_0
   \zeta_0+\zeta_0-1))+ \nonumber\\
   &&+4 K_1 \Omega_{\Lambda} \Omega_M ((9 \Omega_M-8) T_0^2+(8-9 \Omega_M)
   T_0-2)+ \nonumber\\
   &&+3 \Omega_{\Lambda} \Omega_M (-4 K_2 (T_0-1) T_0+9 \Omega_M^2-10 \Omega_M)\bigg] \,,
\eea
where we used the Einstein equation at the center $(\eta=\eta_0,r=0)$
\bea
1&=&\Omega_k(0)+\Omega_M+\Omega_{\Lambda}=-K_0+\Omega_M+\Omega_{\Lambda},\\
\Omega_k(r)&=&-\frac{k(r)}{H_0^2 a_0^2},\\
\Omega_M&=&\frac{\rho_0}{3 H_0^2 a_0^3},\\
\Omega_{\Lambda}&=&\frac{\Lambda}{3 H_0^2}.
\eea
and $T_0=\eta_0(a_0 H_0)$ is determined numerically by imposing the conditions \cite{EneaRomano:2011aa} 
\bea
H^{LTB}&=&\frac{\partial_t a(t,r)}{a(t,r)}=\frac{\partial_{\eta} a(\eta,r)}{a(\eta,r)^2}= (a_0 H_0)\frac{\tilde{a}'(T,r)}{a(T,r)^2}\,, \label{HLTB}\\
a(\eta_0,0)&=&a_0 \,, \label{a0}\\
H^{LTB}(\eta_0,0)&=&H_0 \label{H0}\,.
\eea

Finally we can observe that all the above formulae reduce to the well known FLRW form in the homogeneous limit  limit, i.e. when $\{K_1=K_2=0\}$.

\section{Testing the accuracy of the formula}
In order to verify the accuracy of the formula obtained we consider the example of an inhomogeneity described by 
\bea
K(r)&=&\epsilon(1+r+r^2) \,,
\eea
where $\epsilon$ is parametrizing the deviation from a homogeneous cosmological model.
We then compute the corresponding luminosity distance by integrating numerically the Einstein's equations and the geodesic equations,
and compare the numerical results to the red-shift expansion for different values of $\epsilon$.
\begin{figure}
	\centering
		\includegraphics{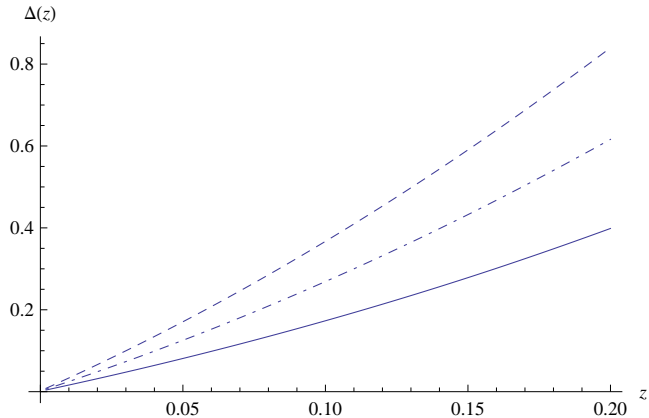}
	\caption{The percentual error $\Delta=100 \frac{ D^{LTB}_{num}-D^{LTB}_{Taylor}}{D^{LTB}_{num}}$ between the numerically computed $D^{LTB}_{Num}(z)$ and the Taylor third order expansion $D^{LTB}_{Taylor}(z)$  is plotted as  a function of the redshift for the LTB solution corresponding to $K(r)=\epsilon(1+r+r^2)$. The solid line corresponds to $\epsilon=0.075$, the dot-dashed line to $\epsilon=0.05$ and the dashed line to $\epsilon=0.1$. }
	\label{fig:Deltaz}	
\end{figure}

As it can be seen in the figure the formula is quite accurate up to a red-shift of $0.2$, where according to the value of $\epsilon$ the percentual error is approximately between $0.3\%$ and $0.7\%$.
The one provided here is only an example to give a preliminary test of the accuracy of the formula, and as such it does not have any direct connection with the actual size of an inhomogeneity which may be surrounding us.
We will investigate more extensively in a separate upcoming paper the range of applicability of the formula in relation with observational data fitting.
We also report the percentual error of the formula for the radial coordinate $r(z)$ as a function of the redshift.
It is important to observe that $r(z)$ depends on our choice of coordinates, which is $\rho_0(r)=const. $, but it is still useful to check its accuracy since it is used in the derivation of the formula for the luminosity distance.
This latter one is a physical observable and so its relation with the red-shift is independent of our coordinate choice except for the fact that coefficients of the Taylor expansion of $K(r)$ would change if we would choose another coordinate system.

\begin{figure}
	\centering
		\includegraphics{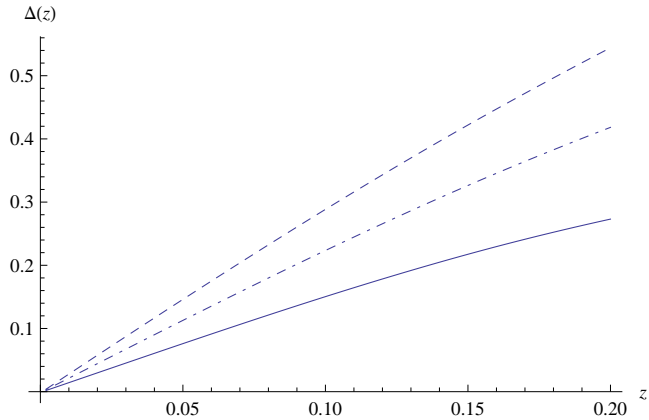}
	\caption{The percentual error $\Delta=100 \frac{ r^{LTB}_{num}-r^{LTB}_{Taylor}}{r^{LTB}_{num}}$ between the numerically computed $r^{LTB}_{Num}(z)$ and the Taylor third order expansion $r^{LTB}_{Taylor}(z)$  is plotted as  a function of the redshift for the LTB solution corresponding to $K(r)=\epsilon(1+r+r^2)$. The solid line corresponds to $\epsilon=0.075$, the dot-dashed line to $\epsilon=0.05$ and the dashed line to $\epsilon=0.1$. }
	\label{fig:Deltaz}
\end{figure}

\section{Conclusion}
We have derived the analytical low red-shift expansion of the luminosity distance for a central observer at the center of a spherically symmetric matter inhomogeneity in presence of a cosmological constant.
We have first solved the null radial geodesic equation and calculated the local red-shift for $r(z)$ and $\eta(z)$, and we have then used these to calculate the expansion of the luminosity distance. The formulae obtained take a simpler form in the case in which $K_0=0$, while in general are rather long and complicated, but can be reduced to a more tractable form in the limit in which the deviation form homogeneity can be treated perturbatively.

The formulas we have derived can be used to understand the physical effects of local inhomogeneities in presence of a cosmological constant. It has the advantage, contrary to previous numerical studies, of not depending on any functional ansatz for the profile of the local inhomogeneity.
This makes it particularly useful to study possible low red-shift inhomogeneities in a model independent way in the regime in which
perturbation theory cannot be applied.

\appendix
\section{Derivation of the analytical formulae}

In order to obtain the formula for $D_L(z)$ in the form in which we reported it in the previous sections we need to apply several simplifying procedures.

The ideas is to express everything in terms of physical quantities, so we can start from the definition of $a_0$ and $H_0$ :
\bea
a_0&=&a(\eta_0,0)=\frac{\rho_0}{k_0+3 \phi_0 } \,,\\
H_0&=&\frac{\dot{a}(t_0,0)}{a(t_0,0)}=-\frac{3 \phi'_0}{2 \rho_0} \,,
\eea
where 
\bea
\phi_0&=&\phi(\eta_0,g_2(0),g_3(0)) \,,\\
\phi'_0&=&\frac{\partial \phi(\eta,g_2(0),g_3(0))}{\partial \eta}|_{\eta=\eta_0}\,,
\eea
and the derivatives respect to the time variable $t$ are obtained in terms of derivatives respect to $\eta$ using equation (\ref{etadef}).
After inverting the above relations we get
\bea
\phi_0=\phi(\eta_0,g_2(0),g_3(0))&=&\frac{\rho_0-a_0 k_0}{3 a_0} \,,\\
\phi'_0=\frac{\phi(\eta,g_2(0),g_3(0))}{d \eta}\bigg|_{\eta=\eta_0}&=&-\frac{2 H_0 \rho_0}{3} \,.
\eea
We can then substitute the above expressions everywhere $\{\phi_0,\phi'_0\}$ appear, which is the reason why they are not present in the formulae obtained.

Another useful relation to simplify intermediate results is the one which can be obtained from the differential equation defining the Weierstrass elliptic function :
\bea
\tilde{\phi}'_0&=&\sqrt{-\frac{K_0^3}{216}-\frac{K_0^2 \tilde{\phi}_0}{12}+\frac{\Omega_{\Lambda} (\Omega_M^0)^2}{16}+4 \tilde{\phi}_0^3}\,.
\eea
It can also be shown that the above relation is equivalent to the Einstein's equation at the center $(\eta_0,0)$
\bea
1&=&-K_0+\Omega_{\Lambda}+\Omega_M \,,
\eea
since solving the Einstein's equation is reduced to solving the Weierstrass equation by construction \cite{EneaRomano:2011aa}.
\section{General formulae and perturbative limit}
In this appendix we give the formulae when $K_0$ is not zero.

For $\eta(z)$ and $r(z)$ we have
\bea
r_2 &=& \frac{1}{1944 a_0 H_0 \Omega_{\Lambda} \Omega_{M}^2 (4 K_0^3-27 \Omega_{\Lambda}\Omega_{M}^2)}
\bigg[-128 K_0^6 K_1 (T_0-1) T_0^2+64 K_0^5 K_1 T_0 ((3 \Omega_{M}-2) \nonumber \\ 
  & &  T_0^2+(2-3 \Omega_M) T_0-2)-972 K_0^4 \Omega_{\Lambda} \Omega_M^2 (K_1 T_0^3-4)+108
   K_0^3 \Omega_{\Lambda} \Omega_M^2 (27 \Omega_M (K_1 T_0^3-2) \nonumber\\
	&&-K_1 T_0 (11
   T_0^2+T_0 (54 \zeta_0+34)-18))-27 K_0^2 K_1 \Omega_{\Lambda} \Omega_M^2 ((81
   \Omega_M^2-66 \Omega_M+8) T_0^3- \nonumber\\
	&& 8 T_0^2 (\Omega_M (81 \zeta_0+39)-81 \zeta_0-17)+4 T_0 (135
   \Omega_M-108 \zeta_0-34)+\nonumber\\
	&&144 (3 \zeta_0+1))-1458 K_0 \Omega_{\Lambda} \Omega_{M}^2 (K_1 (2 \Omega_{M}
   (T_0^2 (6 \Omega_{\Lambda}-9 \zeta_0-1)+ \nonumber\\
	&&T_0 (-6 \Omega_{\Lambda}+6 \zeta_0+4)-6 \zeta_0+2)+3 \Omega_{M}^2
   T_0 (3 T_0 \zeta_0+T_0-4)+8 (T_0^2-T_0+1) \zeta_0)+18 \Omega_{\Lambda}
   \Omega_{M}^2)+ \nonumber\\
	&&4374 \Omega_{\Lambda}^2 \Omega_{M}^3 (K_1 ((6 \Omega_{M}-4) T_0^2+(4-6 \Omega_{M})
   T_0-4)+9 \Omega_{M}^2\bigg]\,,
\eea
\bea
\eta_1 &=& \frac{1}{{972 a_0 H_0 \Omega_{\Lambda} \Omega_M^2 (4 K_0^3-27 \Omega_{\Lambda} \Omega_M^2)}}\bigg[64 K_0^5 K_1 (T_0-1) T_0^2+486 K_0^3 \Omega_{\Lambda} \Omega_M^2 (K_1
   T_0^3-8)+  \\
	& & -27 K_0^2 K_1 \Omega_{\Lambda} \Omega_M^2 T_0 ((27 \Omega_M-4) T_0^2-4 T_0
   (27 \zeta_0+17)+72)+\nonumber \\
	&&-1458 K_0 K_1 \Omega_{\Lambda} \Omega_M^2 T_0 (\Omega_M (3 T_0
   \zeta_0+T_0-4)-4 (T_0-1) \zeta_0)+8748 \Omega_{\Lambda}^2 \Omega_M^3 (K_1 (T_0-1) T_0+3
   \Omega_M)\bigg] \,,\nonumber
\eea
After substituting in the formula for the luminosity distance we get
\bea
D^{\Lambda LTB}_2 &=&\frac{1}{{972 a_0 H_0 \Omega_{\Lambda} \Omega_M^2 (4 K_0^3-27 \Omega_{\Lambda} \Omega_M^2)}}\bigg[64 K_0^5 K_1 (T_0-1) T_0^2+486 K_0^3 \Omega_{\Lambda} \Omega_M^2 (K_1
   T_0^3-8)+\nonumber \\
&&-27 K_0^2 K_1 \Omega_{\Lambda} \Omega_M^2 T_0 ((27 \Omega_M-4) T_0^2-4 T_0
   (27 \zeta_0+17)+72)+\nonumber \\
&&-1458 K_0 K_1 \Omega_{\Lambda} \Omega_M^2 T_0 (\Omega_M (3 T_0
   \zeta_0+T_0-4)-4 (T_0-1) \zeta_0)+8748 \Omega_{\Lambda}^2 \Omega_M^3 (K_1 (T_0-1) T_0+\nonumber \\
&&3
   \Omega_M)\bigg]\,.
\eea
We do not report higher order terms because the expressions are extremely long and would not add any physical insight, but we consider the case in which we can treat perturbatively the function $K(r)\propto \epsilon$, where $\epsilon$ stands for a small deviation from a flat $\Lambda CDM$ model. 

In this perturbative limit for $\eta(z)$ and $r(z)$ we get:
\bea
\eta_1&=&-\frac{1}{a_0 H_0}+\frac{(K_1 T_0-K_1 T_0^2) \epsilon }{3 a_0 H_0 \Omega_M}+O[\epsilon ]^2 \,,\\
\eta_2&=&
\frac{3 \Omega_M}{4 a_0 H_0}+\frac{1}{12 a_0 H_0 \Omega_M}\bigg[-4 K_1-6 K_0 \Omega_M+4 K_1 T_0+4 K_2 T_0-9 K_1 \Omega_M T_0-4 K_1 T_0^2+\nonumber \\
&&-4 K_2 T_0^2+9 K_1 \Omega_M T_0^2) \epsilon \bigg]+O[\epsilon ]^2\,,
\eea
\bea
r_2&=&
-\frac{3 \Omega_M}{4 (a_0 H_0)}+\frac{(2 K_1+3 K_0 \Omega_M-2 K_1 T_0+3 K_1 \Omega_M T_0+2 K_1 T_0^2-3 K_1 \Omega_M T_0^2) \epsilon }{6 a_0 H_0 \Omega_M}+O[\epsilon ]^2 \,,\\
r_3&=&
\frac{-4 \Omega_M+9 \Omega_M^2}{8 a_0 H_0}+\frac{1}{6 a_0 H_0 \Omega_M}\bigg[(2 K_2-5 K_1 \Omega_M-9 K_0 \Omega_M^2-2 K_2T_0+8 K_1 \Omega_M T_0+3 K_2 \Omega_M T_0+\nonumber \\
&& -9 K_1 \Omega_M^2 T_0+2 K_2 T_0^2-8 K_1 \Omega_M T_0^2-3 K_2 \Omega_M T_0^2+9 K_1 \Omega_M^2 T_0^2) \epsilon\bigg]
+O[\epsilon ]^2\,,
\eea
Finally substituting in $D_L(z)$ the above expansion we obtain: 
\bea
D^{\Lambda LTB}_2&=&
(\frac{1}{H_0}-\frac{3 \Omega_M}{4 H_0})+\frac{(K_0+K_1 T_0-K_1 T_0^2) \epsilon }{2 H_0}+O[\epsilon ]^2 \,,\\
D^{\Lambda LTB}_3&=&
\frac{-10 \Omega_M+9 \Omega_M^2}{8 H_0}+\frac{1}{6 H_0}\bigg[3 K_0-2 K_1-9 K_0 \Omega_M+8 K_1 T_0+3 K_2 T_0-9 K_1 \Omega_M T_0-8 K_1 T_0^2+\nonumber \\ 
  & &-3 K_2 T_0^2+9 K_1 \Omega_M T_0^2) \epsilon \bigg]+O[\epsilon ]^2\,.
\eea
As it can be seen $\zeta_0$ is not present in the first order perturbative corrections to a homogeneous universe.

\begin{acknowledgments}
Chen and Romano are supported by the Taiwan NSC under Project No.\
NSC97-2112-M-002-026-MY3, by Taiwan's National Center for
Theoretical Sciences (NCTS). Chen is also supported by the US Department of Energy
under Contract No.\ DE-AC03-76SF00515.
AER is also supported by UDEA under the programs GFIF Sostenibilidad, dedicacion exclusica and the CODI project IN10219CE.

\end{acknowledgments}

\end{document}